\documentstyle[amssymb,aps]{revtex}

\begin{document}
\title{Finite-temperature scalar fields and the cosmological constant in an
Einstein universe}
\author{M. B. Altaie$^{1}\thanks{%
Electronic mail: maltaie@yu.edu.jo}$ \ and M.R. Setare$^{2,3,4}\thanks{%
Electronic mail: rezakord@yahoo.com}$}
\author{$^{1}$Department of Physics, Yarmouk university, 21163 Irbid-Jordan}
\author{$^{2}${Institute for Theoretical Physics and Mathematics, Tehran, Iran }}
\author{$^{3}${Department of Science, Physics group, Kordestan University, Sanandeg,
Iran}}
\author{$^{4}${Department of Physics, Sharif University of Technology, Tehran, Iran }}
\date{December 2002}
\maketitle

\begin{abstract}
We study the back reaction effect of massless minimally couupled scalar
field at finite temperatures in the background of Einstein universe.
Substituting for the vacuum expectation value of the components of the
energy-momentum tensor on the RHS of the Einstein equation, we deduce a
relationship between the radius of the universe and its temperature. This
relationship exhibit a maximum temperature, below the Planck scale, at which
the system changes its behaviour drastically. The results are compared with
the case of a conformally coupled field. An investigation into the values of
the cosmological constant exhibit a remarkable difference between the
conformally coupled case and the minimally coupled one.
\end{abstract}

\section{Introduction}

Quantum fields in curved spacetimes have been investigated by many authors
(for a thorough in-depth review see\cite{Birrel}). The basic question was:
does the zero-point energy of a quantized field act as the source of a
gravitational field? The works dealing with this question started by mid
1970's when matter fields were brought into connection with spacetime
curvature through the calculation of the vacuum expectation value of the
energy-momentum tensor $<0|T_{\mu \nu }|0>$ \ [2-6].\thinspace The
motivations for studying this quantity stems from the fact that $T_{\mu \nu
} $ is a local quantity that can be defined at a specific spacetime point,
contrary to the particle concept which is global. The energy-momentum tensor
also act as a source of gravity in the Einstein field equations, therefore
it is expected that $<0|T_{\mu \nu }|0>$ can plays an important role in any
attempt to model a self-consistent dynamics involving the classical
gravitational field coupled to the quantized matter fields. So, once $%
<0|T_{\mu \nu }|0>$ is calculated in a specified background geometry, we can
substitute it on the RHS of the Einstein field equation and demand self
consistency. For models with a cosmological constant like the Einstein
static model this means that 
\begin{equation}
R_{\mu \nu }-\frac{1}{2}g_{\mu \nu }R+g_{\mu \nu }\Lambda =-8\pi <0|T_{\mu
\nu }|0>,  \label{q1}
\end{equation}
where $R_{\mu \nu }$ is the Ricci tensor, $g_{\mu \nu }$ is the metric
tensor and $R$ is the scalar curvature.

The solution of (\ref{q1}) will determine the development of the spacetime
in presence of the given matter field, for which $|0>$ can be defined
unambiguously. This is known as the ''back-reaction problem''. It is
interesting to perform the calculation of $<0|T_{\mu \nu }|0>$ in
Friedman-Robertson-Walker (FRW) models since it is believed that the real
universe is, more or less, a sophisticated form of the Friedman models.
However the time-dependence of the spacetime metric generally creates
unsolvable fundamental problems. One such a problem was the definition of
vacuum in a time-dependent background \cite{Ful1}; a time-dependent
background is eligible for producing particles continuously, therefore, pure
vacuum states in the Minkowskian sense do not exist. On the other hand, an
investigation into the thermodynamics of a time-dependent systems lacks the
proper definition of thermal equilibrium, which is a basic necessity for
studying finite-temperature field theory in curved backgrounds \cite{Ken}.

During the last twenty five years numerous papers have been published
dealing with quantum field theoretic calculations performed in curved
spacetimes, and specifically in the Einstein universe. On basic reason fore
choosing the Einstein universe is that it stands the two fundamental
challenges mentioned above. Being static, the Einstein universe leaves no
ambiguity in defining the vacuum both locally and globally \cite{Birrel}.
This same feature also allows for thermal equilibrium to be defined
unambiguously. Ford\cite{Ford1} has shown that a closed Robertson-Walker
universe has the same vacuum energy and pressure as a static universe of
instantaneously equal radius. Furthermore, the Einstein static metric can be
related to the closed Robertson-Walker metric through the conformal
transformations, further it was shown by Kennedy \cite{Ken} that thermal
Green functions for the static Einstein universe and the time-dependent
Robertson-Walker universe are conformally related, hence deducing a (1-1)
correspondence between the vacuum and the many particle states of both
universes. So that, under the equilibrium condition, the thermodynamics of
quantum fields in an Einstein universe of radius $a$ is equivalent to that
of an instantaneously static closed Friedman-Roberson-Walker (FRW) universe
of equal radius [2,5,9]. This means that the results obtained in a closed
FRW universe would be qualitatively the same as those obtained in an
Einstein universe. This means that successive states of the Einstein
universe can represent a working model that will include features of the FRW
universe with non-zero cosmological constant if no geometrodynamical effect
is assumed to exist.

The finite-temperature corrections to the vacuum energy of the universe is
an important factor in defining its thermal development. Thus it is
interesting to consider such calculation, and it would be of further
interest to consider the back reaction of such cases on the development of
the very early universe.

Altaie and Dowker \cite{AD} calculated the finite temperature corrections to
the massless conformally coupled scalar field, the neutrino field and, the
photon field in the background of an Einstein universe. The results of the
calculation for the photon field was used to deduce a self-consistent
solution for the Einstein field equation, i.e., a back-reaction problem,
from which a relation between the temperature and the radius of the Einstein
universe was deduced. However, this relation was not fully exploited at that
time and therefore some of the thermodynamical aspects were kept unexposed.

Recently \cite{Altaie1}(hereafter will be referred to as I), we have
investigated the back reaction effects of the conformally coupled scalar
field and the photon fields in the background of the Einstein universe. We
solved the Einstein field equation in each case and found a relation between
the temperature and the radius of the universe. This relation exhibited a
minimum radius below which no self-consistent solution for the Einstein
field equation can be found. Also we have found that the system exhibit two
different behaviors, one at very small radii where the temperature rises
sharply with the radius until it reaches a maximum value, after which it
decreases as the radius is increased. We called the first regime as the
''Casimir regime'' since the vacuum energy is dominant through it and we
called the next one the ''Planck regime'' since the energy of the system
follows a Planckian distribution.

In this paper we will consider the calculation of the back reaction effects
of the minimally coupled massless scalar field at finite temperatures in the
background of the Einstein static universe. The aim is to expose the thermal
behavior of the system and compare the results with those of the conformally
coupled case. Taking into consideration the renewed interest in the
cosmological constant, we will also investigate the relationship between the
values of the cosmological constant and the temperature for successive
states of the Einstein universe for both the conformally coupled and the
minimally coupled cases, in order to expose the different roles played by
both fields during the Casimir and the Planck regimes. This consideration
will shed light on the question of the decay of the cosmological constant
during early stages of the universe (corresponding to small radii states of
\ Einstein universe), such a question comes in the context of inflationary
cosmology. Results shows that there is qualitative and an important
difference in the behavior of the cosmological constant for both fields in
the Casimir regime. Throughout this paper we use the natural units in which $%
G=c=${\it 
h\hskip-.2em\llap{\protect\rule[1.1ex]{.325em}{.1ex}}\hskip.2em%
}=$k_{B}=1$.

\section{Basic formalism}

The metric of the static Einstein universe is given by 
\begin{equation}
ds^{2}=dt^{2}-a^{2}\left[ d\chi^{2}+\sin^{2}\chi\left( d\theta^{2}+\sin
^{2}\theta d\phi^{2}\right) \right] \,,  \label{q2}
\end{equation}
where $a$ is the radius of the spatial part of the universe $S^{3}$ and, $%
0\leq\chi\leq\pi$, $0\leq\theta\leq\pi$, and $0\leq\phi\leq2\pi$.

We consider an Einstein universe being filled with a massless quantum gas in
thermal equilibrium at temperature $T$. The total energy density of the
system can be written as

\begin{equation}
<T_{00}>_{tot}=<T_{00}>_{T}+<T_{00}>_{0},  \label{q3}
\end{equation}
where $<T_{00}>_{0}$ is the zero-temperature vacuum energy density (Casimir
energy density) and $<T_{00}>_{T}\,$is the corrections for finite
temperatures, i.e.,

\begin{equation}
<T_{00}>_{T}=\frac{1}{V}\sum_{n}\frac{d_{n}\epsilon _{n}}{\exp \beta
\epsilon _{n}-1},  \label{q4}
\end{equation}
where $\epsilon _{n}$ and $d_{n}$ are the eigen energies and degeneracies of
the $n$th state, and $V$=$2\pi ^{2}a^{3}$ is the volume of the spatial
section of the Einstein universe.

Normally $<T_{00}>_{T}$is divided into two parts: the black body term ,
which is calculated by converting the summation into an integration and is
denoted by $<T_{00}>_{T}^{b}$, and the correction term, calculated from the
remainder and denoted by $<T_{00}>_{T}^{a}$ (see \cite{AD}), so that (\ref
{q3}) can be written as

\begin{equation}
<T_{00}>_{tot}=<T_{00}>_{T}^{b}+<T_{00}>_{T}^{a}+<T_{00}>_{0}.  \label{q5}
\end{equation}

Experience tells us that in the limit $\xi \equiv Ta\rightarrow 0,$we have
(see\cite{AD})

\begin{equation}
\lim\limits_{\xi \rightarrow 0}<T_{00}>_{T}^{a}=-<T_{00}>_{T}^{b},
\end{equation}
so that at the low temperature limit we are left with the Casimir term only.
Whereas in the high-temperature limit we have

\begin{equation}
\lim\limits_{\xi \rightarrow \infty }<T_{00}>_{T}^{a}=-<T_{00}>_{0},
\label{q7}
\end{equation}
so that we are left with the black body term only. In the next section we
will use these limits to obtain the value of the Casimir energy and the
expression for the black body term.

To investigate the back-reaction effect of finite-temperature quantum fields
on the behavior of the spacetime we should substitute for $<T_{00}>_{tot}$
on the RHS of the Einstein field, but this time with the cosmological
constant $\Lambda $, i.e.

\begin{equation}
R_{\mu \nu }-\frac{1}{2}g_{\mu \nu }R+g_{\mu \nu }\Lambda =-8\pi <T_{\mu \nu
}>_{tot}.  \label{q8}
\end{equation}

All the Einstein field equations for the system are satisfied due to the
symmetry of the Einstein universe which is topologically described by $%
T\otimes S^{3}$, and due to the structure of $<T_{\mu\nu}>$ in this geometry
which comes to be diagonal, and is given by (see [1],p.186 )

\begin{equation}
<T_{\mu }^{\nu }>=\frac{p(s)}{2\pi ^{2}a^{4}}diag(1,-1/3,-1/3,-1/3),
\label{q9}
\end{equation}
where $p(s)$ is a spin-dependent coefficient.

Since we are interested in the energy density, we will consider the $T_{00}$
only. In order to eliminate $\Lambda $ from (\ref{q8}) we multiply both
sides with $g_{\mu \nu }$ and sum over $\mu $ and $\nu $, then using the
fact that $T_{\mu }^{\mu }=0$ for massless fields, and for the Einstein
universe $R_{00}=0$ , $g_{00}=1$ ,and $R=\frac{6}{a^{2}}$, we get

\begin{equation}
\frac{6}{a^{2}}=32\pi <T_{00}>_{tot}.  \label{q10}
\end{equation}

Note that in the general case conformal anomalies do appear in the
expression for $<T_{\mu}^{\mu}>$ , but because of the high symmetry enjoyed
by the Einstein universe these anomalies do not appear and $<T_{\mu}^{\mu}>$
is found to be traceless for massless particles.

\section{The vacuum energy and back reaction}

The minimally coupled massless scalar field satisfies the covariant
Klein-Gordon equation

\begin{equation}
\square \phi =0,  \label{q11}
\end{equation}

where $\square=\nabla_{\mu}\nabla^{\mu}.$

Eq. (\ref{q11}) was solved by Schrodinger\cite{Schr} for the case of closed
universe, the energy eigenvalues are given by

\begin{equation}
\epsilon _{n}=\frac{[n(n+2)]^{1/2}}{a},\text{ \ \ \ \ }n=0,1,2,3...
\label{q12}
\end{equation}

and the degeneracy of each energy level is $d_{n}=(n+1)^{2}.$

For a minimally coupled massless scalar field in an Einstein universe the
total energy density is therefore given by

\begin{equation}
<T_{00}>_{tot}=\frac{1}{2\pi ^{2}a^{4}}\sum_{n=0}^{\infty }\frac{%
(n+1)^{2}[n(n+2)]^{1/2}}{\exp ([n(n+2)]^{1/2}/\xi )-1}+<T_{00}>_{0},
\label{q13}
\end{equation}

In the low-temperature limit (or small radius) we find that

\begin{equation}
\lim\limits_{\xi \rightarrow 0}<T_{00}>_{tot}=0,
\end{equation}
which, by (\ref{q5}) and (\ref{q7}), means that the renormalized vacuum
energy density for the minimally coupled scalar field at zero temperature
vanish, i.e.

\begin{equation}
<T_{00}>_{0}=0,  \label{q15}
\end{equation}

\bigskip Indeed this result can be confirmed by applying the Abel-Plana
summation formula

\begin{equation}
\sum_{n=1}^{\infty }f(n)=\int_{0}^{\infty }f(x)dx-\frac{1}{2}%
f(0)+i\int_{0}^{\infty }\frac{f(ix)-f(-ix)}{e^{2\pi x}-1}dx
\end{equation}

directly to the energy mode-sum.

In the high-temperature limit we deduce that

\begin{equation}
<T_{00}>_{T}^{b}=\frac{\pi ^{2}}{30}T^{4}+\frac{1}{12}\frac{T^{2}}{a^{2}},
\label{q16}
\end{equation}
which is just half the value obtained for the photon field as would be
expected. However in the limit of very large radius $a$ we obtain the usual
value of the back body radiation term, i.e.

\begin{equation}
\mathrel{\mathop{\lim }\limits_{\xi \longrightarrow \infty }}%
<T_{00}>_{tot}=\frac{\pi ^{2}}{30}T^{4}  \label{q17}
\end{equation}

In order to investigate the back-reaction effect of the field we substitute
for $<T_{00}>_{tot}$ from (\ref{q13}) in (\ref{q10}) and request a
self-consistent solution, we get

\begin{equation}
a^{2}=\frac{8}{3\pi }\sum_{n=0}^{\infty }\frac{(n+1)^{2}[n(n+2)]^{1/2}}{\exp
([n(n+2)]^{1/2}/\xi )-1}  \label{q18}
\end{equation}

This equation determines a relation between the temperature $T$ and the
radius $a$ of the Einstein universe in presence of the minimally coupled
massless scalar field. The solutions of this equation are shown in Fig. 1 in
comparison with the results obtain earlier for the conformally coupled case.
Here again two regimes are recognized: one corresponding to small values of $%
\xi $ where the temperature rises sharply reaching a maximum at $%
T_{max}\approx 0.6T_{p}=0.85\times 10^{32}$K. Since this regime is
controlled by the vacuum energy (the Casimir energy), therefore we call it
the ''Casimir regime''. The second regime is what we call the ''Planck
regime'', which correspond to large values of $\xi $, and in which the
temperature asymptotically approaches zero for very large values of $a$.

From (\ref{q10}) and (\ref{q17}) we can calculate the background (Tolman)
temperature of the universe in the limit of high temperature and large
radius. This is found to be

\begin{equation}
T_{b}=\left( \frac{45}{8\pi ^{3}a^{2}}\right) ^{1/4},  \label{q19}
\end{equation}
This is the same result we obtained for the conformally coupled case
discussed in I. To get a glimpse of the meaning of this result, we may
substitute for $a$ the present value of the Hubble length, i.e. $%
a=1.38\times 10^{28}$ cm we obtain $T=31.556$ K. Conversely if we demand
that the background temperature have the same value as the present
equivalent temperature of the CMB radiation, i.e. $2.73$ K, then the radius
of the Einstein universe should be $1.\,29\times 10^{30}cm$. This is about
two orders of magnitude larger than the estimated Hubble length.

The reason for the coincidence of the behavior of the minimally coupled and
conformally coupled scalar fields in the Planck regimes stems from the fact
that the field equations differs only by a factor of $1/a^{2}$ which becomes
arbitrarily small for large values of $a.$ This means that the difference
between the behaviors of the two fields can only be noticeable within the
Casimir regime, and this difference will become even clearer in the next
section when we consider the cosmological constant. In the massive cases
this factor can be absorbed into the mass itself and consequently one can
differentiate between the two field at the very early stages of the universe
only, but as the radius of the universe grows large the difference between
the minimally coupled scalar field and the conformally coupled one becomes
undetectable. This conclusion is quite general and would apply in the case
of FRW universe too.

\section{The cosmological constant}

The cosmological constant was first introduced by Einstein in order to
justify the equilibrium of a static universe against its own gravitational
attraction. The discovery of Hubble that the universe may be expanding led
Einstein to abandon the idea of a static universe and, along with it the
cosmological constant. However the Einstein static universe remained to be
of interest to theoreticians since it provided a useful model to achieve
better understanding of the interplay of spacetime curvature and of quantum
field theoretic effects. Recent year have witnessed a resurgence of interest
in the possibility that a positive cosmological constant $\Lambda $ may
dominate the total energy density in the universe (for recent reviews see 
\cite{Carrol} and \cite{Sahni}). At a theoretical level $\Lambda $ is
predicted to arise out of the zero-point quantum vacuum fluctuations of the
fundamental quantum fields. Using parameters arising in the electroweak
theory results in a value of the vacuum energy density $\rho _{vac}=10^{6}$
GeV$^{4}$ which is almost $10^{53}$ times larger than the current
observational upper limit on $\Lambda $ which is $10^{-47}$ GeV$^{4}\sim
10^{-29}$ gm/cm$^{3}$. On the other hand the QCD vacuum is expected to
generate a cosmological constant of the order of $10^{-3}$ GeV$^{4}$ which
is many orders of magnitude larger than the observed value. This is known as
the old cosmological constant problem. The new cosmological problem is to
understand why $\rho _{vac}$ is not only small but also, as the current
observations seem to indicate, is of the same order of magnitude as the
present mass density of the universe.

The value of the cosmological constant for an Einstein universe seem to be
trivial. It is directly related with the total energy density. However,
since the energy density in an Einstein universe varies inversely with $%
a^{2} $ and not with $a^{3}$, new features are expected in the behavior of
the cosmological constant. In what follows we are going to investigate the
possible values of the cosmological constant for different radii of the
Einstein universe in presence of the massless conformally coupled and
minimally coupled scalar field. But since different radii of the universe
corresponds to different temperature with a non-trivial relationship between
the radius and the temperature as was found in sec. II of this paper, the
values of the cosmological constant at different temperatures turns out to
be non-trivial and is rather of some serious interest as we find a
qualitative differences between the two types of fields.

Contracting the field equations in (\ref{q8}) we find that

\begin{equation}
\Lambda =\frac{R}{4}=\frac{3}{2a^{2}}.  \label{q20}
\end{equation}

On the other hand the Einstein field equations reduces to

\begin{equation}
-\frac{3}{a^{2}}+\Lambda =-8\pi \rho _{tot},  \label{q21}
\end{equation}

and

\begin{equation}
-\frac{1}{a^{2}}+\Lambda =\frac{8\pi \rho _{tot}}{3}  \label{q22}
\end{equation}

where $\rho_{tot}=<T_{0}^{0}>_{tot}$. Solving the above two equations we
obtain

\begin{equation}
\Lambda =8\pi \rho _{tot}  \label{q23}
\end{equation}

Here we will consider $\rho _{tot}=\rho _{vac}+\rho _{rad},$ but in a more
general case one can set $\rho _{tot}=\rho _{vac}+\rho _{rad}+\rho _{matter}$%
, with $\rho _{rad}$ belonging to the massless field filling the spatial
part of the universe and $\rho _{matter}$ belonging to the pressureless dust
that may exist. The addition of the energy density of the pressureless
matter will not make any qualitative change in the results since $\rho
_{matter}$ in an Einstein universe specifically behaves same as $\rho _{vac}$
and $\rho _{rad}.$

Using (\ref{q20}) and the results obtained in the previous section for the
dependence of $T$ on $a$ we can solve for the dependence of $\Lambda $ on $T$%
. Fig. (2) depicts the relationship between the cosmological constant $%
\Lambda $ and the temperature for successive states of the Einstein universe
under the effect of back reaction of the minimally coupled scalar field at
finite temperatures in comparison with the conformally coupled field . It
shows that the cosmological constant for the minimally coupled case decays
monotonically from an infinite value all through the vacuum dominated regime
until it reaches a comparatively small value at a critical temperature at
which the system changes its behavior into the Planck regime. Whereas the
case of conformally coupled case we notice that the value of $\Lambda $ is
nearly constant throughout the Casimir regime and only start decaying in the
Planck regime. From the point of view of inflationary models a large value
of $\Lambda $ is needed to resolve the problem of horizon and the problem of
flatness, and possibly to generate seed fluctuations for galaxy formation 
\cite{Sahni}.

\section{Discussion and conclusions}

One of the interesting points of this paper is the clear difference between
the behavior of the conformally coupled massless scalar field and the
minimally coupled scalar field as a result of the back reaction, this
behavior which becomes clearer when the value of the cosmological constant
is considered. However, one may feel uneasy with the energy scale that
appears to be higher than the Planck scale in the case of the conformally
coupled scalar field which was considered in I. In fact this is only
fictituous because if we consider the collective effect of many fields then
the resultant effect will bring the energy scale below the Planck energy.
This can be easily checked if one solves the Einstein field equation with
both the minimally coupled and the conformally coupled energies added up in
the source, which confirms the conclusion in \cite{Kaufman} that the range
of validity of the quasi-classical approximation can be extended for large
number of fields.

The conformal relationship between the static Einstein universe and the
Robertson-Walker universe and the possibility to consider the Einstein
universe of a given radius as representative of an instantaneously static
Robertson-Walker universe \cite{Ford1} and the (1-1) correspondence between
the vacuum and the many particle states of both universes as established by
the work of Kennedy \cite{Ken}, suggests that the thermal behavior of a real
closed universe is qualitatively similar to the results obtained in this
work. Therefore, we feel that the calculations in the Einstein universe are
useful in understanding the interplay between quantum fields and the
curvature. Indeed in I our calculations showed that an Einstein universe
with curvature radius of about two order of magnitude larger than the Hubble
radius will have the same CMB temperature as the presently measured one. On
the other hand the analysis of the most recent observations of CMB spectrum
suggests that the curvature radius of the real universe is at least 50 times
larger than the Hubble radius \cite{Turner}. This is a point in favour of
the practical relevance of the calculations.

The main findings are same as previously stated in I but with an additional
point concerning the cosmological constant:

1. The thermal development of the universe is a direct consequence of the
state of its global curvature.

2. Unless containing some pressureless matter the Einstein universe will be
singular at zero temperature in presence of the minimally coupled massless
scalar field, in contrast with the conformally coupled case where a non-zero
radius was found as an effect of the back reaction of the non-zero Casimir
energy. A non-zero expectation value of the vacuum energy density always
implies a symmetry breaking event.

3. During the Casimir regime the universe is totally controlled by vacuum.
The energy content of the universe is a function of its radius. Using the
conformal relation between the static Einstein universe and the closed FRW
universe \cite{Ken}, this result indicates that in a FRW\ model there would
be a continuous creation of energy out of vacuum as long as the universe is
expanding, a result which was confirmed by Parker long ago \cite{Parker1}.

4. The cosmological constant arising from the minimal scalar field mostly
decays during the Casimir regime, whereas the conformally coupled scalar
field dominates most of the Casimir regime and part of the Planck regime.
This indicates that the minimally coupled scalar field plays most of its
effective role in regions of high curvatures.

Recenly it was shown by Ellis and Maartens \cite{Ellis} that the birth of an
inflating universe from the state of a static Einstein universe containing
minimally coupled scalar field and ordinary matter, is quite possible under
certain conditions. Such models are shown to avoid the quantum gravity era.
However, as remarked by the authors, the fine tuning problem in these models
is to fix the initial radius. Therefore, the results presented in this paper
may find some applications in such an approach for closed inflationary
cosmologies.

\begin{center}
------------------------------------------------------------------
\end{center}

\bigskip {\bf Figure Caption}

FIG 1. Comparison between the temperature-radius relationship for massless
conformally coupled (dashed line) and the minimally coupled (solid line)
scalar fields.

FIG 2. Comparison between the contributions of the massless conformally
coupled (dashed line) and the minimally coupled (solid line) scalar fields
to the cosmological constant in an Einstein universe at finite temperatures.

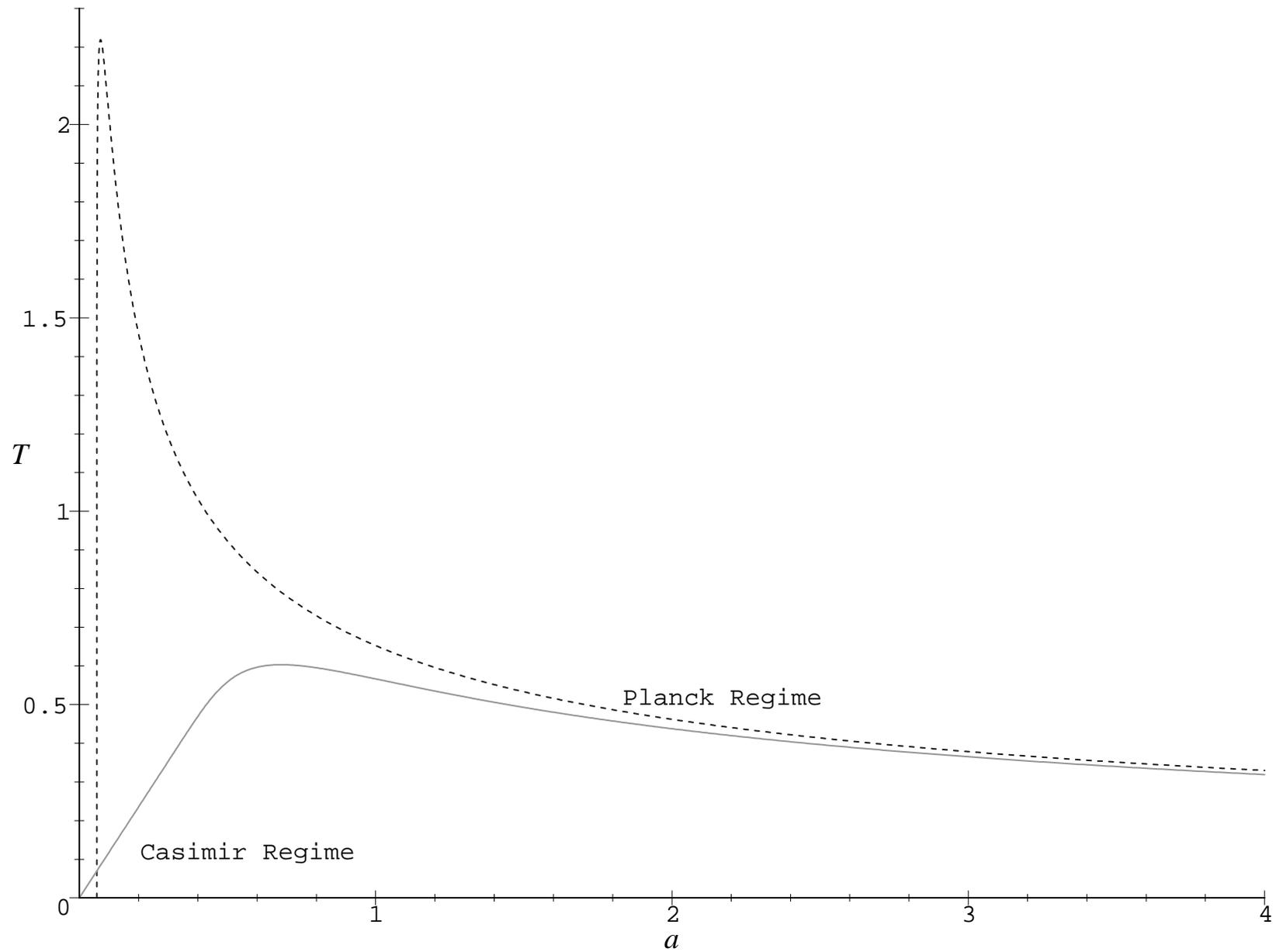
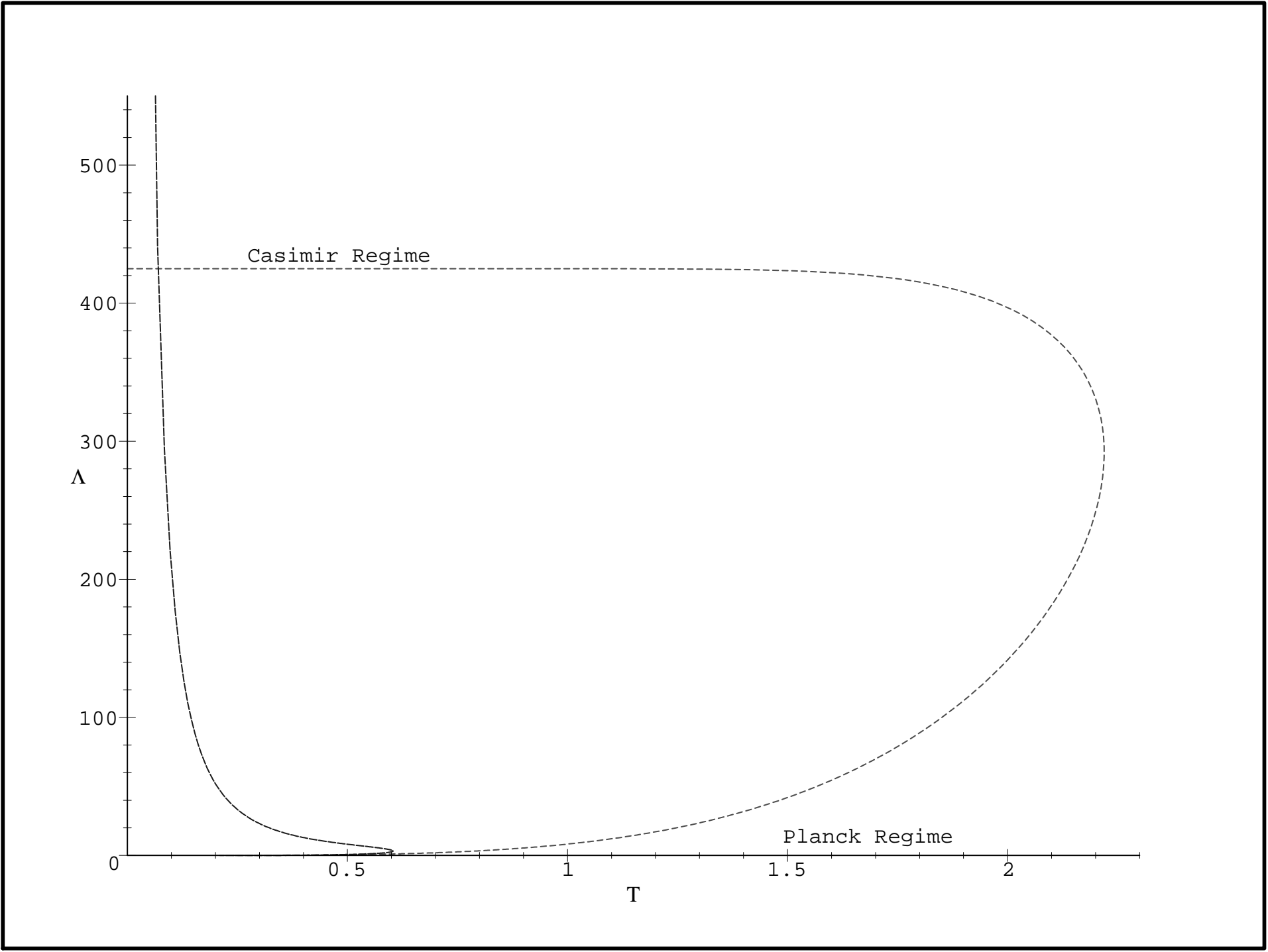
\end{document}